# THE ENVIRONMENT OF "E+A" GALAXIES

Ann I. Zabludoff[1], Dennis Zaritsky[2], Huan Lin[3], Douglas Tucker[4], Yasuhiro Hashimoto[5], Steven A. Shectman[1], Augustus Oemler[5], and Robert P. Kirshner[6]

[1]Observatories of the Carnegie Institution of Washington, 813 Santa Barbara St., Pasadena, CA 91101, E-mail: aiz@ociw.edu, shec@ociw.edu

[2]UCO/Lick Observatory and Board of Astronomy and Astrophysics, University of California at Santa Cruz, Santa Cruz, CA, 95064, E-mail: dennis@ucolick.org

[3]Department of Astronomy, University of Toronto, Toronto, Ontario, M5S 1A7, Canada, E-mail: lin@astro.utoronto.ca

[4]Astrophys. Inst. Potsdam, An der Sternwarte 16, D-14482, Potsdam, Germany, E-mail: dlt@aip.de

[5]Astronomy Department, Yale University, New Haven, CT, 06520, E-mail: oemler@astro.yale.edu, hashimot@sparx.astro.yale.edu

[6]Astronomy Department, Harvard University, Cambridge, MA, 02138, E-mail: kirshner@cfa.harvard.edu

## ABSTRACT

The spectrum of an "E+A" galaxy (Dressler & Gunn 1983), which is dominated by a young stellar component but lacks the emission lines characteristic of any significant, on-going star formation, suggests that the galaxy experienced a brief, powerful starburst within the last Gyr (Dressler & Gunn 1983; Couch & Sharples 1987). In past work, this violent star formation history and the detection of these galaxies almost exclusively in distant clusters linked them to the "Butcher-Oemler effect" (Butcher & Oemler 1978) and argued for the influence of cluster environment in the evolution of galaxies. However, no statistical survey of the environments of "E+A"s had ever been made. From 11113 galaxy spectra in the Las Campanas Redshift Survey (Shectman *et al.* 1992), we have obtained a unique and well-defined sample of 21 nearby "E+A" galaxies with the same spectral characteristics as "E+A"s in distant clusters. These "E+A"s are selected to have the strongest Balmer absorption lines (the average equivalent width of H$\beta$, $\gamma$, $\delta$ is $> 5.5$ Å) and weakest [O II] emission-line equivalent widths ($< 2.5$ Å, which corresponds to a detection of [O II] of less than $2\sigma$ significance) of any of the galaxies in the survey. In contrast to inferences drawn from previous studies, *we find that a large fraction ($\sim$ 75%) of nearby "E+A"s lie in the field, well outside of clusters and rich groups of galaxies.* We conclude that interactions with the cluster environment, in the form of the intracluster medium or cluster potential, are not essential for "E+A" formation and therefore that the presence of these galaxies in distant clusters does not provide strong evidence for the effects of cluster environment on galaxy evolution.

If one mechanism is responsible for "E+A" formation, then the observations that "E+A"s exist in the field and that at least five of the 21 in our sample have clear tidal features argue that galaxy-galaxy interactions and mergers are that mechanism. The most likely environments for such interactions are poor groups of galaxies, which have lower velocity dispersions than clusters and higher galaxy densities than the field.

Groups are correlated with rich clusters and, in hierarchical models, fall into clusters in greater numbers at intermediate redshifts than they do today (cf. Bower 1991; Lacey & Cole 1993; Kauffmann 1994). When combined with the strong evolution observed in the field population (cf. Broadhurst *et al.* 1988; Lilly *et al.* 1995), our work suggests that the B-O effect may reflect the typical evolution of galaxies in groups and in the field rather than the influence of clusters on the star formation history of galaxies.

## 1. Introduction

Little is known about how environment influences the evolution of galaxies. Although we can divide the environmental factors that may transform galaxies into two broad categories, 1) interactions and mergers between galaxies and 2) interactions between a galaxy and the gravitational potential or hot intracluster medium of a cluster of galaxies, differentiating between galaxy-galaxy and galaxy-cluster mechanisms is difficult in practice. Extensive observations exist to support the claim that galaxy-galaxy interactions increase star formation rates (Lonsdale, Persson, & Matthews 1984; Kennicutt *et al.* 1987; Sanders *et al.* 1988). However, the effects of the cluster environment on galaxy evolution are harder to isolate. Differences between the morphologies (Dressler 1980), spectral properties (Rose *et al.* 1994), HI extents (Giovanelli and Haynes 1985; Cayatte *et al.* 1994), and chemical abundances (Skillman *et al.* 1995) of cluster and field galaxies could reflect environment-dependent galaxy *formation* rather than environment-dependent evolution. Discerning the influence of cluster-galaxy interactions on galaxy evolution in clusters is further complicated by the infall of galaxies from the field. Therefore, much effort has been made to identify an evolutionary sequence that is found only in the cluster environment.

Two key observations have been used to argue for the existence of a cluster-driven evolutionary sequence. First, some intermediate redshift clusters ($z \sim 0.3$) contain a much larger fraction of blue, actively star forming galaxies than do nearby clusters ("the Butcher-Oemler effect"; Butcher & Oemler 1978 (BO78)). Second, "E+A" galaxies (Dressler & Gunn 1983 (DG83)), whose strong Balmer absorption but lack of strong [O II] emission suggest that they are evolved Butcher-Oemler type galaxies in which the starburst (or period of elevated star formation) ended within roughly the last Gyr, appear to be absent from the intermediate redshift field. These observations have fueled speculation about cluster-specific mechanisms — such as ram pressure stripping by the intracluster medium (DG83) or interaction with the cluster potential (Byrd and Valtonen 1990) — that might have been more effective at earlier times and thus responsible for the apparent rapid evolution of cluster galaxies.

Yet, the argument for cluster-driven galaxy evolution remains uncertain. Recent work has shown an enhancement in the number of actively star forming galaxies in the intermediate redshift *field* (cf. Broadhurst, Ellis, & Shanks 1988; Glazebrook *et al.* 1995; Lilly *et al.* 1995) implying that the Butcher-Oemler (B-O) effect might result more from a universal change in galaxy properties than from evolving cluster environments. The statistics of these distant samples are not yet sufficient to determine whether B-O galaxies are an intrinsically different population than these active field galaxies or whether the fraction of B-O galaxies in clusters is significantly higher than that of the active population in the field.

The question of whether "E+A" galaxies form only in clusters and are thus direct evidence for a star formation trigger exclusive to the cluster environment is also unresolved, not only because the galaxy statistics of intermediate redshift field surveys are much poorer than for similarly distant clusters, but also because "E+A"s are only a small fraction ($\sim$ 10%; DG83; Fabricant, McClintock, & Bautz 1991 (FMB91)) of the detected intermediate redshift cluster populations. However, there are some recent clues to the answer. First, one nearby "E+A" galaxy (G515; Oegerle, Hill, & Hoessel 1991) does not seem to lie in a rich cluster and has a tidal feature suggesting an encounter with another galaxy. Second, some luminous merging galaxies in the field, with strong Balmer absorption and moderate [O II] $<$ 15 Å emission, are likely "E+A" progenitors (Liu & Kennicutt 1995). These objects are possible precursors to an "E+A" field population and a connection to G515. Third, a recent study by Caldwell *et al.* (1993 (CRSEB93)) found a sample of 12 galaxies in (or projected on) the Coma cluster with enhanced Balmer absorption and no detectable emission lines. This last result indicates that although nearby "E+A" galaxies are scarcer than those at intermediate redshifts (from $<$ 1% at $z \sim 0$ to $\sim$ 10% at $z > 0.3$; DG83; FMB91), "E+A"s exist in sufficiently large numbers locally to allow us to classify their environments.

To resolve the controversy about where and how "E+A"s form, and to examine one of the strongest pieces of evidence for cluster-driven galaxy evolution, we have undertaken a extensive search for nearby "E+A" galaxies using the Las Campanas Redshift Survey (LCRS; Shectman *et al.* 1995b). With its high quality spectra and dense sampling of galaxies, the LCRS is ideal for comparing uniformly-selected, statistical samples of galaxy populations and their environments. For the first time, we can quantify the unusual spectral characteristics of "E+A" galaxies in relation to a representative sample of $\sim 11000$ nearby galaxies ($0.05 < z < 0.13$) and compare the distributions of local density for "E+A" and non-"E+A" galaxies.

In this paper, we present a well-defined sample of 21 "E+A" galaxies with high signal-to-noise spectra ($S/N > 8$). These spectra will serve as templates for future studies of "E+A"s in nearby clusters and at higher redshifts. With these data, we can determine if "E+A"s frequent the nearby field, and if, by extension, influences other than cluster environment are responsible for their formation.

This paper is organized as follows. Section 2 describes the LCRS data from which we draw the "E+A" sample and the "E+A" sample selection criteria. The spectra and the distribution of galaxy luminosities are discussed in §3. In §3, we also calculate the distributions of local densities for "E+A" and non-"E+A" galaxies and examine the morphologies of the "E+A"s. Section 4 is a discussion of why the data argue that the most probable mechanism for "E+A" formation is galaxy-galaxy interactions and mergers. We summarize our conclusions in §5.

## 2. The Data

### 2.1. The LCRS Spectra

For our spectroscopic sample we use the data from the LCRS (Shectman *et al.* 1992, 1995ab). The fiber spectra were obtained with the multi-fiber spectrograph (Shectman *et al.* 1992) and Reticon detector mounted on the DuPont 2.5m telescope at the Las Campanas Observatory. Each spectrum is extracted from the two-dimensional array, flat fielded, wavelength calibrated, and finally sky subtracted based on the flux normalization of the 5577 Å night sky line. The spectra have a resolution of approximately 5 Å, a pixel scale of $\sim$ 3 Å, and a wavelength range of 3500-6500 Å. The average signal-to-noise ($S/N$) in the continuum around the H$\beta$, H$\gamma$, and H$\delta$ absorption lines is typically 8 to 9. The $S/N$ is calculated by determining the ratio of the mean square deviation about the continuum near each absorption line (not including the absorption line or any nearby sky lines) and the mean continuum at those three positions.

The current LCRS consists of 23700 galaxy spectra with a mean redshift of $z \sim 0.1$ in six slices of $1.5^\circ \times 80^\circ$. Three of the slices are in the northern galactic hemisphere ($\alpha \sim 10^h$-$15.3^h$ for $\delta = -3^\circ$, $-6^\circ$, and $-12^\circ$, respectively) and three are in the southern hemisphere ($\alpha \sim 21^h$-$4.6^h$ for $\delta = -39^\circ$, $-42^\circ$, and $-45^\circ$, respectively). A slice is comprised of 20-30 $1.5^\circ \times 1.5^\circ$ fields, each containing $\sim$ 100 galaxies with Gunn r-band isophotal magnitudes between $15.0 \leq m_r \leq 17.7$ and with surface brightnesses within the 3.5 arcsec fiber aperture of $\mu_{a,r} \lesssim 21$ mag arcsec$^{-2}$ (cf. Tucker 1994). (Note that $\sim$ 20% of the galaxies have somewhat different selection criteria, because they lie in the earliest survey fields in which $\sim$ 50, not $\sim$ 100, of brightest galaxies were observed (cf. Shectman 1995ab; Lin *et al.* 1995 for further details).) The survey fields are uniformly sampled and are 70% complete on average. In fields where there are more possible targets than fibers, the selection of targets satisfying the magnitude and surface brightness criteria is random. Galaxies brighter than $m_r = 15.0$ are excluded to avoid having a few very bright objects dominate the telescope pointing during the spectroscopic exposure, because the pointing and field rotation are adjusted to maximize the total count rate. The shape of the luminosity function of the LCRS is consistent (Lin *et al.* 1995) with that of other redshift surveys (Loveday *et al.* 1992; Marzke *et al.* 1995), and it is therefore unlikely that the applied magnitude and surface brightness cuts significantly bias the LCRS sample relative to other surveys.

### 2.2. Selection of the "E+A" Sample

"E+A" galaxies have distinctive spectra characterized by strong $H\delta$ $\lambda$4102, $H\gamma$ $\lambda$4340, and $H\beta$ $\lambda$4861 absorption lines *and* little or no [O II] $\lambda$3727 emission (DG83). This combination of features is rare because the presence of a young "A" stellar component, which gives rise to the Balmer lines, shows that there has been significant recent star formation, but the lack of [O II] emission indicates that there is little or no current star formation. The combination of spectral properties is frequently interpreted as evidence for a starburst that ended within roughly the last Gyr (DG83; Couch & Sharples 1987 (CS87)). The detection of metallic absorption lines, such as Mg *b* $\lambda$5175, Ca H & K $\lambda$3934, 3968, and Fe $\lambda$5270, indicate that there is an additional old or "E" (for "elliptical"; DG83) stellar population in these galaxies that is mostly G, K, and M stars. The "E" designation came about because "E+A"s were spectroscopically discovered, but the name is somewhat misleading because there is no conclusive evidence that these objects are exclusively spheroidal. In fact, some "E+A"s are disk-like (Franx 1993, Wirth, Koo, & Kron 1994, Dressler *et al.* 1994; Couch *et al.* 1994; Caldwell *et al.* 1995 (CRFL95)) and thus, as pointed out by Franx, "E+A"s are best described neutrally as "K+A"s. CS87 further divide the galaxies with strong Balmer absorption and weak [OII] emission into two subclasses, which may have different evolutionary histories: blue post-starburst (PSG) and redder $H\delta$-strong (HDS) objects (also see FMB91; §4 of this paper). In this paper, our use of the term "E+A" includes the full range of "E+A" morphology and colors, and we postpone a discussion of the different subclasses of "E+A"s until §4. Finally, we stress that our galaxies are different than those with strong Balmer absorption lines, but with significant [O II] emission (cf. Newberry, Boroson, & Kirshner 1990), which are sometimes also referred to as "E+A" galaxies and are a component of the active population in the distant field.

We begin our search for "E+A" galaxies by setting our spectroscopic selection criteria. We set fairly conservative limits on the redshift range and line strengths of the objects in order to select the highest quality "E+A" sample, a luxury afforded us by the large size of the LCRS survey. First, we select only those galaxies in the LCRS with recessional velocities between 15000 and 40000 km s$^{-1}$ for two reasons: (1) to utilize the relatively constant selection function of the LCRS over this range (Tucker 1994; Lin *et al.* 1995) and (2) to reduce the aperture bias caused by the 3.5 arcsec aperture of the fibers on galaxies of large angular extent (cf. Zaritsky, Zabludoff, & Willick 1995). Between 15000 and 40000 km s$^{-1}$, the 3.5 arcsec fiber aperture subtends projected diameters between 2.3 kpc and 5.3 kpc ($q_0 = 0.5$ and $H_0 = 100$ km s$^{-1}$ Mpc$^{-1}$ are used throughout this paper). Second, we exclude low S/N spectra because they complicate the equivalent width measurements. Specifically, we exclude galaxies whose spectra have an average signal-to-noise ratio $S/N \leq 8$ in the continua about the $H\delta$, $H\gamma$, and $H\beta$ lines (the average $S/N$ of the galaxies with velocities between 15000 and 40000 km s$^{-1}$ is $\sim 10$). The $S/N$ cut excludes $\sim 50\%$ of the galaxies with $17.0 < m_r < 17.7$, but does not significantly alter the number of brighter galaxies in the sample. After the redshift and $S/N$ cuts, the LCRS sample consists of 11113 galaxies.

We automate the calculation of the S/N, Balmer line ($H\delta$, $H\gamma$, and $H\beta$) equivalent widths, the [O II]$\lambda$3727 equivalent width and uncalibrated flux, and the 4000 Å break ($D_{4000}$). Because the spectra are not flux calibrated, only relative spectral measurements over narrow wavelength ranges are possible (*i.e.* no absolute fluxes or colors can be measured). To calculate an equivalent width, the algorithm fits the local continua over the 100 pixels ($\sim$ 250Å) on either side of the line that excludes the line itself and nearby sky lines. Beginning at the line center, the algorithm integrates the line outward until reaching the continuum level. That uncalibrated flux and the interpolated value of the continuum at line center are used to calculate the equivalent width. Equivalent widths are cosmologically corrected. Finally, because the CH G-band at 4304 Å interferes with the continuum measurement immediately to the blue of $H\gamma$, we integrate the flux of the H$\gamma$ line redward of line center and double that value. The average equivalent width of the three measured Balmer absorption lines is denoted $\langle H \rangle$. The equivalent width uncertainties, which are typically less than 1 Å, are calculated using counting statistics (the detector is a photon counter with approximately zero read noise), the local noise in the continuum, and standard propagation of errors.

The 4000 Å break amplitude ($\mathrm{D}_{4000}$), a measure of the composition of the stellar population of a galaxy, is calculated by taking the ratio of the mean counts between 4050 Å and 4250 Å to that between 3750 Å and 3950 Å. The narrow range of continuum sampled in the $D_{4000}$ calculation ensures that this differential measure is minimally biased by the unfluxed nature of the LCRS spectra or by reddening. The error in $D_{4000}$ is calculated by propagating the measured noise in the continuum through the calculation.

An "E+A" galaxy lies in the high tail of the distribution of Balmer absorption line strengths and in the low tail of the distribution of [O II] emission. However, there are no distinct and isolated galaxy populations in the $\langle H \rangle$ *vs.* EW[O II] plane (Figure 1). We note that although we, and others, discuss "E+A"s as a distinct population, the definition of these galaxies is somewhat ambiguous and arbitrary — "E+A" galaxies are only one extreme of a continuum of properties. Most galaxies in the nearby universe have 0 Å $< \langle H \rangle <$ 3 Å and EW[O II] $<$ 10 Å. The most active galaxies, which have the strongest [O II] and Balmer line *emission*, occupy the lower right corner of the plane in Figure 1. In contrast, galaxies with the "E+A" characteristics of strong Balmer *absorption* and little or no [O II] emission occupy the extreme upper left. It is from this region that we draw our "E+A" sample.

As mentioned previously, the galaxy statistics and the $S/N$ of the spectra in the LCRS sample allow us to define an "E+A" sample somewhat more conservatively than for samples at higher redshifts, where estimates of $\langle H \rangle$ and EW[O II] are more uncertain. Typically, the equivalent widths of the higher redshift "E+A"s are $\langle H \rangle > 5$ Å and EW[O II] $< 5$-10 Å (DG83; CS87; FMB91). We choose a lower limit of $\langle H \rangle = 5.5$ Å for what we identify as an "E+A" . We further select only those galaxies with EW[O II] $< 2.5$ Å, which for this sample corresponds to an emission line that is detected at less than $2\sigma$ significance (Figure 1 inset). The sample of 21 galaxies that satisfy these criteria is listed in Table 1 and totals 0.2% of the LCRS galaxies with $S/N > 8$ and recessional velocities between 15000 and 40000 km s$^{-1}$.

Table 1 lists the galaxy name, 1950 coordinates, apparent r-band magnitude ($m_r$), heliocentric velocity ($cz$), absolute magnitude ($M_r$) corrected for cosmology and galactic extinction where possible (from the measurements of Burstein & Heiles (1982)), EW[O II] and its error, $\langle H \rangle$ and its error, $D_{4000}$ and its error, and a flag that indicates if the galaxy lies near a rich group or cluster of galaxies (cf. §3.3).

Given the continuum of properties and the observational uncertainties, our sample of 21 "E+A" galaxies excludes galaxies that might satisfy the selection criteria but which are placed just beyond the selection boundary by observational errors. For example, increasing the EW[O II] limit to 3.5 Å and decreasing the $\langle H \rangle$ limit to 4.5 Å (*i.e.,* including outliers within $\sim 1\sigma$ of the adopted Balmer and [O II] line cutoffs) boosts the "E+A" sample to 67 galaxies, or 0.6% of the LCRS. However, our aim in the current work is not to obtain a complete sample of "E+A"s, but a conservative subsample of galaxies with spectral signatures that are definitely indicative of recent star formation and consistent with those of "E+A"s at higher redshifts.

## 3. Results

### 3.1. Spectral Properties

Even with the strict selection criteria, the spectra of the 21 "E+A" galaxies span a range of spectral characteristics that suggest that these objects are at different evolutionary stages, have experienced different evolutionary histories, and/or had morphologically different progenitors. We number the spectra in order of increasing $D_{4000}$ in Figure 2. The first few spectra in the first panel are dominated by light from a young stellar component. Figure 3 shows one of these extraordinary "A" star-like spectra (#3) with the features labeled (note the absence of [O II] emission). In contrast, the spectra in the last panel of Figure 2 have significantly larger $D_{4000}$, less pronounced Balmer absorption, and stronger Mg $b$ absorption. A detailed discussion of these trends is beyond the scope of this paper and is the topic of a forthcoming paper (Zabludoff *et al.* 1995a) in which we compare spectral population synthesis models (Bruzual & Charlot 1993) with the "E+A" sample.

### 3.2. The Distribution of Luminosities

Determining the distribution of luminosities of nearby "E+A" galaxies is important for understanding if the "E+A" formation mechanism operates on different types of progenitors and for uncovering the possible selection biases in higher redshift "E+A" samples. In intermediate redshift clusters, only the brightest "E+A" galaxies are presently detected, and nothing is known about the existence of fainter "E+A"s. Nearby, where spectroscopic identifications of "E+A"s are easier, there is at least one very luminous ($M_{25,r} = -23.4$) "E+A" galaxy ($z = 0.09$; Oegerle, Hill, & Hoessel 1991) and also several known dwarf "E+A"s ($M \sim -16$; CRSEB93). We can use the "E+A" sample drawn from the LCRS to determine the full distribution of "E+A" luminosities for the first time.

The broad range of r-band luminosities for the "E+A" galaxies in our sample (Figure 4) suggests that "E+A"s evolve from both dwarf and giant progenitors. A Kolmogorov-Smirnov test is unable to distinguish the distribution of r-band absolute magnitudes for the "E+A"s from that of the other LCRS galaxies (the cumulative distributions are only distinguishable at the 7% confidence level). However, this test and its interpretation are somewhat naive — galaxies can brighten significantly in the r-band during a post-starburst phase (Barger *et al.* 1995). In spite of this uncertainty, it is clear that the "E+A" phenomenon is not confined to giant galaxies.

The distribution of $D_{4000}$ (Figure 5) shows that "E+A"s have weaker breaks (younger stellar populations and bluer colors) than the typical LCRS galaxies (the distributions are distinguishable at the $> 99\%$ level). This result is hardly surprising given that we select the "E+A" sample partly on the basis of Balmer absorption line strength. Because "E+A"s have absolute magnitudes consistent with the LCRS as a whole and are generally bluer objects, they have intrinsically brighter blue luminosities than the typical galaxy, as we expect for post-starburst galaxies. This comparatively higher blue luminosity makes "E+A"s somewhat more likely to be detected than intrinsically red cluster galaxies in red-selected surveys of distant clusters ($z > 0.3$; DG83; CS87) in which the blue light is redshifted into the r-band.

### 3.3. The Distribution of Local Densities

The principal aim of this study is to identify the range of environments inhabited by "E+A" galaxies. It is only in low redshift surveys, such as the LCRS, that the spectroscopic sampling of objects is sufficiently dense to confidently ascertain which galaxies are members of rich groups or clusters of galaxies and which lie in the field. To determine the distribution of "E+A" environments, we calculate the local galaxy density for all the galaxies in the LCRS sample. We begin this calculation by counting the total number of galaxies within a square box 1 Mpc wide on the sky and $\pm 1000$ km s$^{-1}$ deep in radial velocity centered on each galaxy. If the central galaxy lies within a projected distance of 0.5 Mpc from a survey boundary, we do not calculate its local density. The total number of LCRS galaxies with $S/N > 8$ and $15000 < cz < 40000$ km s$^{-1}$ for which we determine a local density is 9595 (or 86% of the sample), including 20 of the 21 "E+A"s.

We correct the local density calculation for (1) the incompleteness in the sampling of each fiber field (which varies with the projected density of objects in each field) and (2) the change in the sampling of the luminosity function (LF) of the survey with redshift. For the first correction, we normalize the number of galaxies in the box around each galaxy by the fraction of all galaxies within the survey magnitude limits that have been spectroscopically observed in that field. This correction is typically small due to the high completeness of the survey (cf. §2). For the second correction, we scale the number of galaxy counts in each box by the LF of the survey. The LF is described by a Schechter function (Schechter 1976) of the form $\phi^* = 0.019$, ${M_r}^* = -20.29$, and $\alpha = -0.70$ (Lin *et al.* 1995). Because we know the bright and faint apparent magnitude limits to which each fiber field is sampled and the redshift of the box, we can determine the absolute magnitude limits to which galaxies in the box are sampled (correcting for cosmological effects and for extinction where HI data are available (Burstein & Heiles 1982). (We estimate that the extinction corrections are small ($\lesssim 0.2$ magnitudes) in the few regions where the HI maps are incomplete.) We then scale the observed counts to the number expected to be brighter than a specific absolute magnitude limit. We adopt $M_r = -18.5$ as that magnitude limit as a compromise between maximizing the statistics and limiting the correction factor to a range where the LF is directly measured.

The distribution of scaled galaxy counts (local densities) $N$ for the 9595 galaxies with $S/N > 8$, $15000 < cz < 40000$ km s$^{-1}$, and coordinates well inside the survey boundaries is shown in the bottom panel of Figure 6. The scaled counts around the 20 "E+A"s are in the middle panel. To compare the "E+A" environments with those of galaxies in known clusters and rich groups we plot (top panel) the distribution of local densities for the 320 galaxies with $S/N > 8$ and $15000 < cz < 40000$ km s$^{-1}$ that are classified as members of systems with radial velocity dispersions $\sigma_r > 400$ km s$^{-1}$ in the LCRS group catalog (Tucker *et al.* 1996). The 400 km s$^{-1}$ velocity dispersion lower limit is consistent with the dynamically "coldest" rich groups and clusters (Ramella *et* *al.* 1995). The distributions of local density for the "E+A" galaxies and galaxies in rich systems are distinguishable at the $> 99\%$ confidence level.

Although some of the "E+A"s in our sample obviously lie in lower density regions than does the average galaxy in rich groups and clusters, the three-dimensional, "nearest-neighbor" group identification algorithm (cf. Huchra & Geller 1982) used to compile the LCRS group catalog includes only group members within an isodensity contour of $\Delta\rho/\rho = 80$ (in contrast to $\Delta\rho/\rho = 0$, the definition of the field). As a result, galaxies in the outskirts of rich groups or clusters (at radii of several Mpc where $\Delta\rho/\rho$ declines below 80, but is still well above the field value) are not identified as members of those systems. Therefore, we proceed with a second, more general test to determine whether our "E+A"s lie within an infall radius of the LCRS systems with $\sigma_r > 400$ km s$^{-1}$. We adopt a radius of 5 Mpc, the approximate infall radius for nearby Abell richness class $R \geq 1$ clusters, as a conservative estimate of the infall radius of clusters today.

Three "E+A"s (#20, #11, #4) have a rich group or cluster within 5 Mpc and $\pm 2000$ km s$^{-1}$ (more than 2.5$\times$ the velocity dispersion of a typical Abell $R \geq 1$ cluster (Zabludoff *et al.* 1990)). The other 18 "E+A"s do not lie near hot, dense systems within the survey boundaries. Because the 10 Mpc width of the search box centered on an "E+A" can extend beyond the boundaries of the survey (the LCRS strips are only 1.5° wide in declination, which corresponds to 8 Mpc at our mean redshift, $z \sim 0.1$), some of the sample "E+A"s may lie in the outskirts of rich systems that are beyond the survey region.

To estimate the number of rich systems that may lurk beyond the edges of the survey, we compare the volume of the search boxes that falls within the survey region to the volume that lies outside the survey. We find that 58% of the total volume sampled by the 10 Mpc wide search boxes is within the LCRS. Because we have identified three "E+A" galaxies with rich groups or clusters nearby, it is likely that there are two other rich systems near our "E+A" galaxies that lie outside the survey limits. If we assign these unknown systems to two "E+A" galaxies that appear to be isolated, then we expect that only 5 (± 2 from standard counting statistics) of our sample of 21 "E+A"s are within the infall radius (5 Mpc and ±2000 km $s^{-1}$) of a rich group or cluster.

Therefore, *a significant fraction (∼ 75%) of "E+A" galaxies lie in the nearby field, well outside of the environments of rich groups and clusters.*

Even though the large majority of our "E+A"s are field galaxies, it is still possible that "E+A"s are more likely than other LCRS galaxies to be associated with hot, dense environments. Our sample is insufficiently large to definitively address this issue. A total of 982 of the 11113 LCRS galaxies have a $\sigma_r > 400$ km $s^{-1}$ rich group or cluster within 5 Mpc and ±2000 km $s^{-1}$. The ratio of the survey volume sampled to the total volume sampled by the search boxes is 55%. Therefore, the fraction of LCRS galaxies that are likely to lie near a rich group or cluster, 1785/11113 = 0.16, is not significantly different than the fraction of "E+A"s probably associated with rich systems, 5/21 or $0.24 \pm 0.10$.

### 3.4. "E+A" Morphologies

The results in the previous section demonstrate than "E+A"s form in a range of environments. If one mechanism is responsible for "E+A" formation, that mechanism cannot be one associated exclusively with cluster environment, such as the shocking of a galaxy's interstellar medium by the intracluster medium or the tidal stripping of a galaxy by the global cluster potential. Perhaps interactions or mergers with other galaxies boost the star formation rates of "E+A"s. We can then ask if any of the "E+A"s in our sample exhibit the morphological signatures of galaxy-galaxy encounters, such as tidal tails or companion galaxies. In addressing this question, the advantages of a nearby sample of "E+A" galaxies are obvious.

The morphological and kinematic data for "E+A"s are scarce at present. From the literature, we know only that several "E+A"s are disk galaxies (Franx 1993; Dressler *et al.* 1994; Couch *et al.* 1994; Wirth, Koo, & Kron 1995; CRFL95) and that at least one other "E+A" has a long tidal tail (Oegerle, Hill, & Hoessel 1991). Although we cannot morphologically type the "E+A"s in our sample without deep exposures and detailed kinematic data, digitized sky survey images provide some interesting clues about the formation of these galaxies.

The 21 images in Plate 1 are extracted from scans of SERC $b_J$ or the POSS E plates from the STScI Digitized Sky Survey. Each image has a pixel scale of 1.7 arcsec and is approximately 70 arcsec wide with north at the top and east to the left. The sky in all the images is normalized to the same median and *rms* deviation. The images follow from left to right and downward the sequence of increasing $D_{4000}$ in Figure 2.

Plate 1 shows that the three bluest, A star-dominated galaxies in the sample (#1, #2, and #3) are clearly disturbed. Galaxies #7 and #20 also have tidal features. (We note that galaxy #20, a face-on spiral interacting with a similarly distorted spiral to the south of the image, has the most uncertain "E+A" classification in the sample. The fiber aperture subtends only $\sim$ 3 kpc and thus samples only a small part of the galaxy, whose tidal features extend over $\sim$ 50 kpc in this image and whose integrated spectrum may have emission lines. However, a preliminary analysis of long-slit spectra obtained along the major and minor axes does not reveal any line emission (Zabludoff *et al.* 1995b).) Another six galaxies have tidal features which are marginally visible in the original images, but difficult or impossible to see in this reproduction. For example, the fourth bluest galaxy in the sample (#4) has a thin tidal tail to the east. In all, at least five of the 21 "E+A"s have obvious tidal features, including the three bluest galaxies in the sample, and another six show hints of similar distortions.

In addition to #20, some of the "E+A" galaxies (*e.g.,* #13, #1, #5, #7, and #10) have apparent companions projected within $\sim$ 25-50 kpc. We have not measured redshifts for these galaxies and so are unable to confirm if they are likely to be interacting with the "E+A"s.

## 4. Discussion

Our principal result is that "E+A" galaxies exist in the low redshift field and thus that interactions with the cluster environment, in the form of the cluster potential or the intracluster medium, are not essential for "E+A" formation. If one mechanism is responsible for boosting the star formation in "E+A"s, their existence in the field and the clear tidal features in five of the 21 galaxies argue that galaxy-galaxy interactions or mergers are that mechanism. Additional support for the interaction/merger origin of "E+A" galaxies comes from the recent spectroscopy of known merger remnants (Liu & Kennicutt 1995), whose spectra are consistent with a model in which the starburst is winding down and a dominant "A" stellar component is emerging.

Any viable picture of "E+A" formation must reproduce the three subclasses of "E+A" that have been identified to date: the blue, post-starburst, tidally-disturbed disk galaxies (PSGs; CS87), the redder $H\delta$-strong spheroidals (HDSs; CS87; Couch *et al.* 1994), and the comparably red and $H\delta$-strong disk galaxies in A665 and in Coma (Franx 1993; CRSEB93). These subclasses may reflect differences in the progenitors (disk or spheroidal galaxies), the violence of the interaction (grazing encounter or full-blown merger, between satellite and primary or between two comparably massive galaxies), the time elapsed since the interaction, or possibly the mechanism that triggers star formation (*i.e.,* if there is a mechanism in addition to galaxy-galaxy encounters). CS87 first discussed the division of "E+A" galaxies into blue and red types (also see FMB91), concluding that a substantial starburst is required to model the blue PSGs.

The origin of the redder HDS and Coma "E+A"s is more ambiguous. Originally, HDS and PSG galaxies were thought to be part of a single evolutionary sequence — by comparing their intermediate redshift cluster data with stellar population synthesis models, CS87 concluded that HDS galaxies could evolve from the PSG's (in addition to arising from the truncation of normal star formation in disk galaxies). However, recent work suggests that HDS galaxies are a different population than PSGs, because the HDS galaxies observed in intermediate redshift clusters appear too luminous at 2.2$\mu$m to have evolved from the PSGs in the same cluster (Barger *et al.* 1995). Furthermore, a study of the disk-like "E+A" galaxies in Coma concludes that these objects cannot be modeled as the result of truncated star formation (CRFL95), but are instead consistent with starburst products.

Do our data shed light on any of these issues? The LCRS "E+A" sample, because of our conservative selection criteria, is biased toward PSGs rather than the redder, weaker-lined HDS and Coma-type objects. The case for a merger origin is strongest for PSGs. Not only do some of these objects show tidal features, but their spectra are very similar to those of evolved merger remnants (Liu and Kennicutt 1995), and their large inferred starburst (10% to 100% of the mass; Couch *et al.* 1994; Barger *et al.* 1995; Liu and Green 1995) demands a violent trigger. As discussed above, the origin of the redder, weaker-lined galaxies is less clear. However, like the PSGs, some of these galaxies (*e.g.,* #18 and #19 in our sample) *are in the field*, and thus, at present, no cluster-specific evolutionary process is required to account for any of the subclasses of "E+A"s.

In addition to the observational evidence that "E+A" galaxies are galaxy-galaxy interaction or merger products, numerical simulations of both the dynamical and spectral evolution of interacting disk galaxies support this hypothesis (cf. Mihos & Hernquist 1994). The models are successful at explaining tidal features in detail (cf. Hibbard & Mihos 1995) and predicting a sudden and brief large increase in the star formation rate (Mihos & Hernquist 1994). In such models, close tidal encounters and minor mergers (which involve a satellite or companion with $\sim$ 10% or less of the mass of the primary) enhance the star formation rate but do not destroy the disk of the original spiral galaxy (Walker *et al.* 1995). In contrast, major mergers (in which the masses of the interacting galaxies are comparable) induce substantial starbursts and destroy disks, producing objects with mass profiles similar to that inferred from the optical luminosity profiles of spheroidal galaxies (Mihos & Hernquist 1994). Some merged, spheroidal galaxies may re-acquire their disks at a later time (Hibbard & Mihos 1995). It is still too early to tell if these models, which transform a blue disk galaxy into a red disk or red spheroidal galaxy, can reproduce all three subclasses of "E+A" .

The models fail in so far as that they cannot presently reproduce our bluest galaxies that have $\langle H \rangle \geq 8$ Å and $D_{4000} \leq 1.3$ (Mihos, private communication) and that they contradict the some "E+A" observations (Franx 1993; CRFL95) by predicting that the star formation is concentrated in the nucleus of the galaxy because the gas dissipates efficiently during the merger. Other simulations (Jog & Solomon 1992), in which the merger of two gas rich spirals triggers massive star formation in Giant Molecular Clouds, are somewhat more successful in predicting extended star formation over several kpc.

The statistics of our "E+A" sample prevent us from ascertaining whether "E+A"s are more or less likely to lie in clusters than are other types of galaxies. Therefore, although the simplest hypothesis is still that all "E+A"s form through the mechanism of galaxy-galaxy interactions and mergers, we cannot rule out an enhancement in the past star formation of cluster galaxies relative to the field that would result from an additional, cluster-specific mechanism. If "E+A"s are in the field today and evolve from bursts of star formation, then "E+A"s should be in the field at higher redshifts, $z \sim 0.3$, where starbursts are more common than at the current epoch (Broadhurst, Ellis, & Shanks 1988). The question for future work is whether the fraction of "E+A"s in the intermediate redshift field is consistent with the canonical $\sim 10\%$ that lie in distant clusters.

If "E+A" galaxies do lie in the field at higher redshifts, then neither they nor their likely progenitors, the blue "Butcher-Oemler" actively star forming galaxies, are necessarily tied to the environments of distant clusters. The most likely environments for galaxy-galaxy interactions and mergers are poor groups of galaxies, which have lower velocity dispersions than clusters and higher galaxy densities than the field (cf. Cavaliere *et al.* 1992; Barnes & Hernquist 1992). Poor groups are correlated with rich clusters and, in hierarchical models (Bower 1991; Lacey & Cole 1993; Kauffmann 1994), fall into clusters in greater numbers at intermediate redshifts than they do today. When combined with the strong evolution observed in the field population (cf. Broadhurst *et al.* 1988; Lilly *et al.* 1995), our work suggests that the B-O effect may reflect the evolution of galaxies in certain types of groups and in the field rather than the influence of clusters on the star formation rates of galaxies. Resolving this issue must wait until the completion of large redshift surveys of the distant field and the advent of high-resolution cosmological simulations that include star formation.

What the current models and our data do make clear is that *the "E+A" phase is a signature of the rapid evolution of a galaxy due to a galaxy-galaxy encounter.* Even such violent interactions as major mergers are probably not uncommon in a galaxy's lifetime. For example, in an $\Omega_0 = 1$ universe, hierarchical models predict that 15% of galaxy halos have accreted at least 50% of their mass in major mergers since $z \sim 0.3$ (Carlberg 1990; Toth & Ostriker 1992). An outstanding question for future work is to determine the fraction of galaxies likely to have evolved through an "E+A" phase.

## 5. Conclusions

From 11113 galaxies in the Las Campanas Redshift Survey, the largest imaging and spectroscopic survey of nearby galaxies to date, we have defined a sample of 21 "E+A" galaxies. These galaxies have the strongest Balmer absorption lines and weakest [O II] emission of any galaxies in the survey. "E+A" galaxy spectra are generally interpreted as being a result of a major starburst that ended within the last $\sim$ Gyr.

We find that $\sim$ 75% of "E+A" galaxies lie in the field, well outside of clusters and rich groups. Thus, immersion in the cluster environment is not a necessary condition for "E+A" formation. The detection of "E+A"s in the nearby field suggests that they exist in the field at higher redshifts. This extrapolation implies that the large disparity between the "E+A" populations in nearby and intermediate redshift clusters ($z > 0.3$) may be due in part to the evolution of the "E+A" population outside of hot, dense environments. In other words, the Butcher-Oemler effect might reflect universal galaxy evolution rather than cluster-driven galaxy evolution. At present, it is unclear how the formation of "E+A"s relates to the color evolution of the field with redshift (Lilly *et al.* 1995).

If one mechanism is responsible for the formation of "E+A" galaxies, the most likely mechanism is galaxy-galaxy mergers and interactions. Many "E+A" galaxies not only exist in the field, but have both the spectral and morphological signatures of evolved merger products. Stellar population synthesis models indicate that "E+A" spectra are consistent with post-starburst objects (cf. CS87; CRSEB93), and at least five of the 21 galaxies in our "E+A" sample have tidal features. Because galaxy-galaxy encounters are not rare, an important issue to address in future work is how common the "E+A" phase might be in the evolution of galaxies in general.

We cannot determine from the statistics of the current sample if "E+A" galaxies are more or less likely to lie in clusters than are other LCRS galaxies. Thus we are unable rule out contributions to "E+A" formation and the Butcher-Oemler effect from cluster-specific mechanisms, such as ram pressure stripping (DG83) and interactions with the global potential (Byrd & Valtonen 1990), that may elevate the star formation rate in galaxies and thus the fraction of "E+A"s in clusters. On the other hand, if galaxy-galaxy interactions and mergers are the "E+A" formation mechanism, then the efficiency of this mechanism in poor groups (including galaxy pairs) and the spatial correlation of such groups with rich clusters might also explain an excess of "E+A"s in clusters relative to the field. Furthermore, the higher rate of the infall of these groups into intermediate redshift clusters (cf. Lacey & Cole 1993) may produce the Butcher-Oemler effect. At the present time, there are no compelling data requiring any mechanism other than galaxy-galaxy mergers and interactions to account for "E+A" galaxies and, by association, for the Butcher-Oemler effect in clusters.

We thank Ian Smail for his thorough reading of the manuscript, detailed comments, and helpful insights. Chris Mihos, Julianne Dalcanton, and John Mulchaey also contributed important information. AIZ acknowledges support from the Carnegie and Dudley Observatories and the AAS. DZ acknowledges partial support provided by a California Space Institute grant. The authors acknowledge three NSF grants that have supported the LCRS: AST 87-17207, AST 89-21326, and AST 92-20460.

Table 1: The "E+A" Sample

| Galaxy | $\alpha$ | | | $\delta$ | | | $m_r$ | $cz$ | $M_R$ | [OII] | $<H\delta\gamma\beta>$ | $D_{4000}$ | Cluster? |
|---|---|---|---|---|---|---|---|---|---|---|---|---|---|
| | 1950.0 | | | | | | | km s$^{-1}$ | | EW (Å) | | | |
| 18 | 0 | 20 | 18.82 | −41 | 50 | 15.6 | 16.42 | 17940±36 | −19.65 | 1.75±0.89 | 5.96±0.49 | 1.589±0.027 | N |
| 20 | 0 | 36 | 20.09 | −39 | 13 | 41.5 | 16.29 | 18960±45 | −19.89 | 2.01±0.95 | 5.59±0.98 | 1.720±0.037 | Y |
| 11 | 1 | 12 | 34.57 | −41 | 38 | 21.8 | 17.29 | 36480±44 | −20.14 | 2.16±0.84 | 6.96±0.62 | 1.500±0.032 | Y |
| 9 | 1 | 15 | 24.20 | −41 | 50 | 10.7 | 17.80 | 19530±56 | −18.47 | 1.09±0.75 | 5.73±1.00 | 1.461±0.026 | N |
| 5 | 1 | 56 | 0.12 | −44 | 51 | 49.0 | 17.06 | 35170±31 | −20.25 | 0.38±0.48 | 5.97±0.52 | 1.354±0.023 | N |
| 19 | 2 | 5 | 51.67 | −45 | 35 | 2.8 | 16.75 | 19190±42 | −19.43 | 0.98±0.86 | 6.08±0.84 | 1.648±0.031 | N |
| 10 | 2 | 9 | 44.51 | −44 | 21 | 43.2 | 17.01 | 31460±44 | −20.10 | 0.70±0.90 | 8.02±0.63 | 1.462±0.025 | N |
| 2 | 2 | 15 | 43.24 | −44 | 46 | 36.7 | 16.69 | 29600±42 | −20.31 | 1.25±0.68 | 7.98±0.63 | 1.277±0.019 | N |
| 4 | 3 | 58 | 23.42 | −44 | 43 | 40.3 | 16.01 | 30350±30 | −21.03 | 1.37±0.62 | 9.82±0.68 | 1.346±0.023 | Y |
| 17 | 10 | 11 | 20.17 | −02 | 40 | 53.0 | 17.32 | 18260±49 | −18.79 | 1.68±0.86 | 6.92±0.77 | 1.578±0.024 | N |
| 1 | 10 | 58 | 48.98 | −11 | 54 | 9.8 | 17.38 | 22380±56 | −19.14 | 1.80±0.67 | 8.98±0.79 | 1.153±0.022 | N |
| 21 | 11 | 12 | 52.65 | −06 | 28 | 51.6 | 17.26 | 29810±39 | −19.80 | 0.75±0.90 | 7.47±0.73 | 1.776±0.034 | N |
| 13 | 11 | 17 | 21.50 | −12 | 36 | 13.3 | 16.20 | 28700±38 | −20.81 | 1.63±0.86 | 6.40±0.60 | 1.531±0.028 | N |
| 6 | 11 | 51 | 21.97 | −02 | 53 | 55.1 | 17.14 | 26530±34 | −19.69 | 0.99±0.58 | 7.56±0.60 | 1.372±0.022 | N |
| 12 | 12 | 3 | 25.98 | −02 | 37 | 50.6 | 17.11 | 29120±57 | −19.90 | 1.62±0.86 | 5.86±0.75 | 1.511±0.034 | N |
| 3 | 12 | 6 | 31.34 | −12 | 5 | 55.4 | 15.36 | 24310±39 | −21.38 | −0.29±0.36 | 8.13±0.61 | 1.301±0.022 | N |
| 16 | 12 | 17 | 21.44 | −05 | 57 | 22.7 | 17.02 | 22910±39 | −19.55 | 2.20±0.91 | 6.20±0.65 | 1.554±0.028 | N |
| 14 | 13 | 54 | 20.95 | −12 | 12 | 10.6 | 16.38 | 21120±42 | −20.10 | 0.02±0.34 | 6.32±0.97 | 1.547±0.023 | N |
| 8 | 14 | 29 | 20.26 | −12 | 44 | 18.2 | 17.39 | 33640±46 | −19.99 | 0.84±0.63 | 5.94±0.63 | 1.445±0.026 | N |
| 15 | 14 | 38 | 5.59 | −06 | 27 | 4.9 | 17.52 | 34120±39 | −19.83 | 1.15±0.71 | 6.33±0.58 | 1.548±0.025 | N |
| 7 | 22 | 38 | 17.77 | −38 | 50 | 17.0 | 15.33 | 34220±47 | −21.93 | 0.44±0.58 | 7.28±0.70 | 1.415±0.020 | N |

## Figure Captions

**Figure 1:** Plot of average Balmer line absorption $\langle H \rangle$ *vs.* [O II] line emission EW[O II] for the 11113 LCRS galaxies with $S/N > 8$ and $15000 < cz < 40000$ km s$^{-1}$. The dashed line encloses the region, $\langle H \rangle > 5.5$ Å and EW[O II] $< 2.5$ Å, from which the sample of 21 "E+A" galaxies (large points) is drawn. The inset shows that EW[O II] cut excludes galaxies with a more than $2\sigma$ detection.

**Figure 2:** Spectra of 21 "E+A" galaxies in the LCRS sample numbered in order of increasing 4000 Å break strength $D_{4000}$. The spectra have been de-redshifted to the rest frame and smoothed to the instrument resolution.

**Figure 3:** Identification of lines in the rest frame spectrum of galaxy #3, which is dominated by a young "A" stellar component. The residual sky line at 5577 Å has been excised. Note the absence of [O II] emission.

**Figure 4:** Absolute magnitude $M_r$ distribution of the 11113 LCRS galaxies with $S/N > 8$ and $15000 < cz < 40000$ km s$^{-1}$. The magnitudes have been corrected for extinction (where galactic HI has been measured, Burstein & Heiles 1982) and for cosmological effects. The distribution for the subset of 21 "E+A" galaxies (shaded) is indistinguishable from the distribution of non-"E+A"s. The histograms are normalized so that the area under each curve is 1.

**Figure 5:** Normalized distributions of $D_{4000}$ for the same samples as in Figure 4. The distributions are distinguishable at the $> 99\%$ confidence level, indicating that "E+A" galaxies are bluer than is typical for LCRS galaxies.

**Figure 6:** The distributions of local densities (scaled galaxy counts) for the 9595 galaxies with $S/N > 8$, $15000 < cz < 40000$ km s$^{-1}$, and coordinates well inside the survey boundaries (bottom), the subsample of 20 "E+A" galaxies (middle), and the subsample of 320 galaxies that are classified as members of rich groups or clusters (top). The distributions for the "E+A" galaxies and for the members of rich systems are distinguishable at the $> 99\%$ confidence level.

**Plate 1:** Digitized SERC Sky Survey $b_J$-band and POSS E images from STScI Sky Survey scans of the 21 "E+A" galaxies in order of increasing $D_{4000}$. The size of each image is ∼70 arcsec, and the pixel scale is 1.7 arcsec. Note the clear tidal features in galaxies #1, #2, #3, #7, and #20.

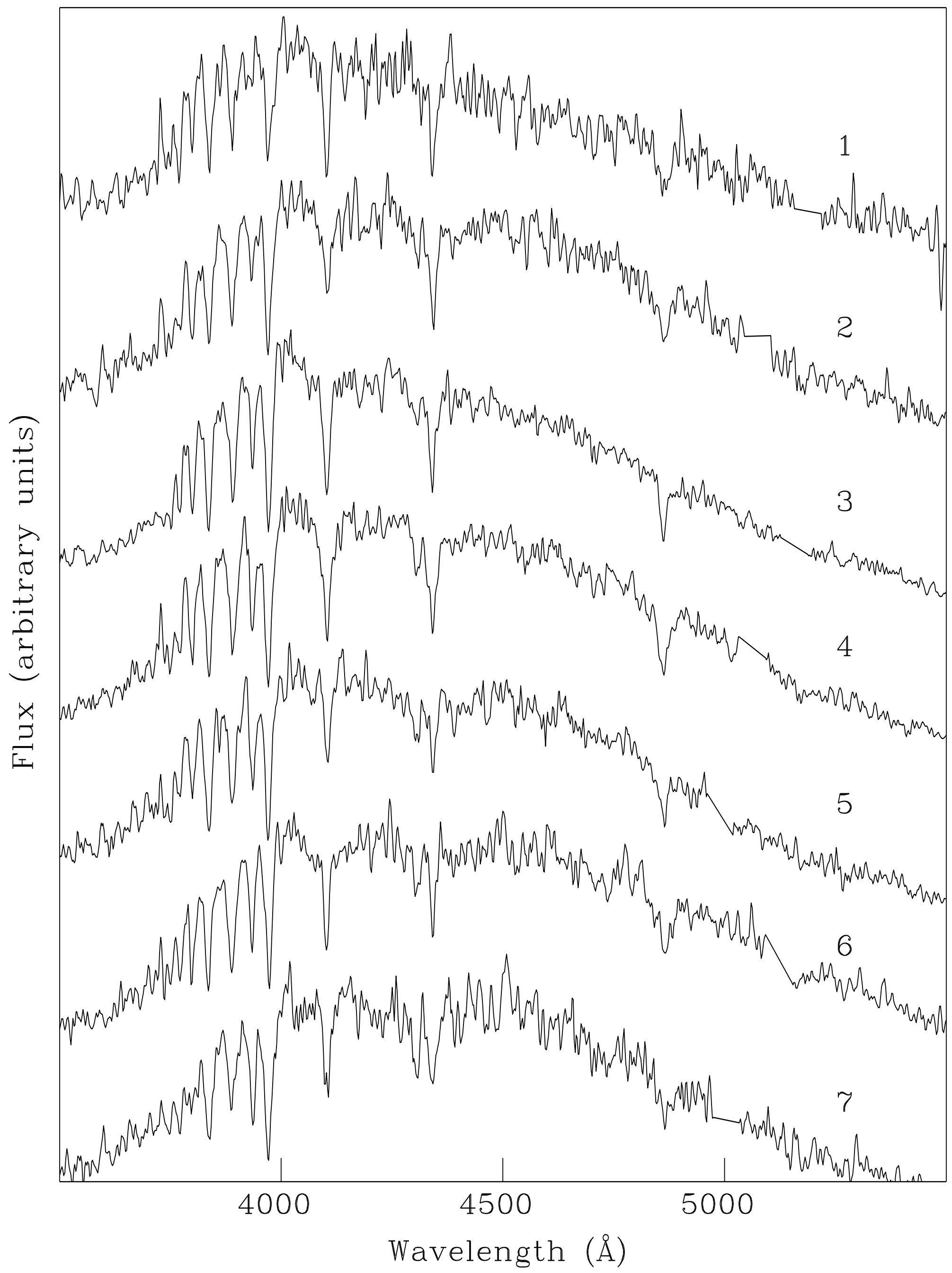

Fig. 1.—

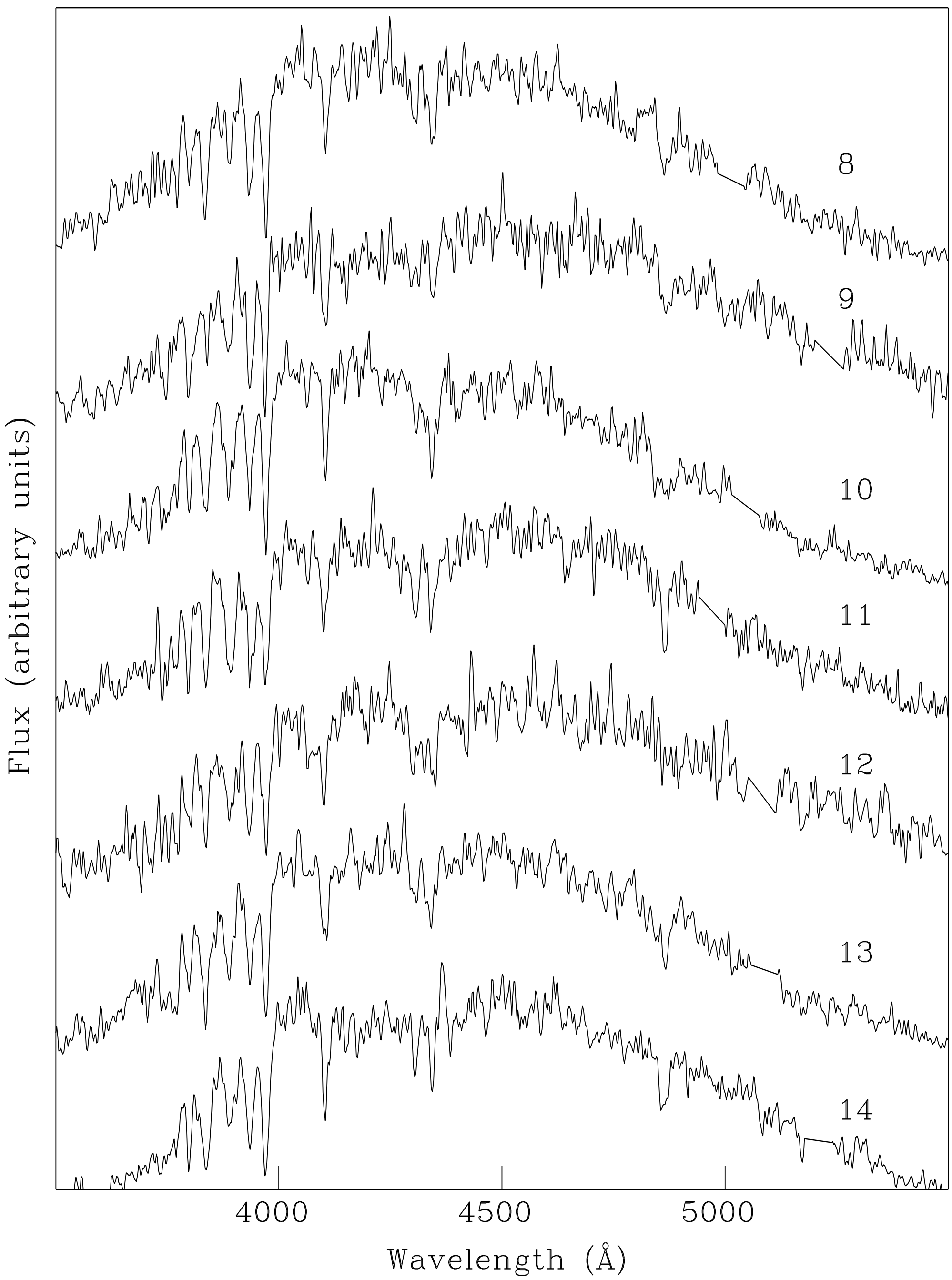

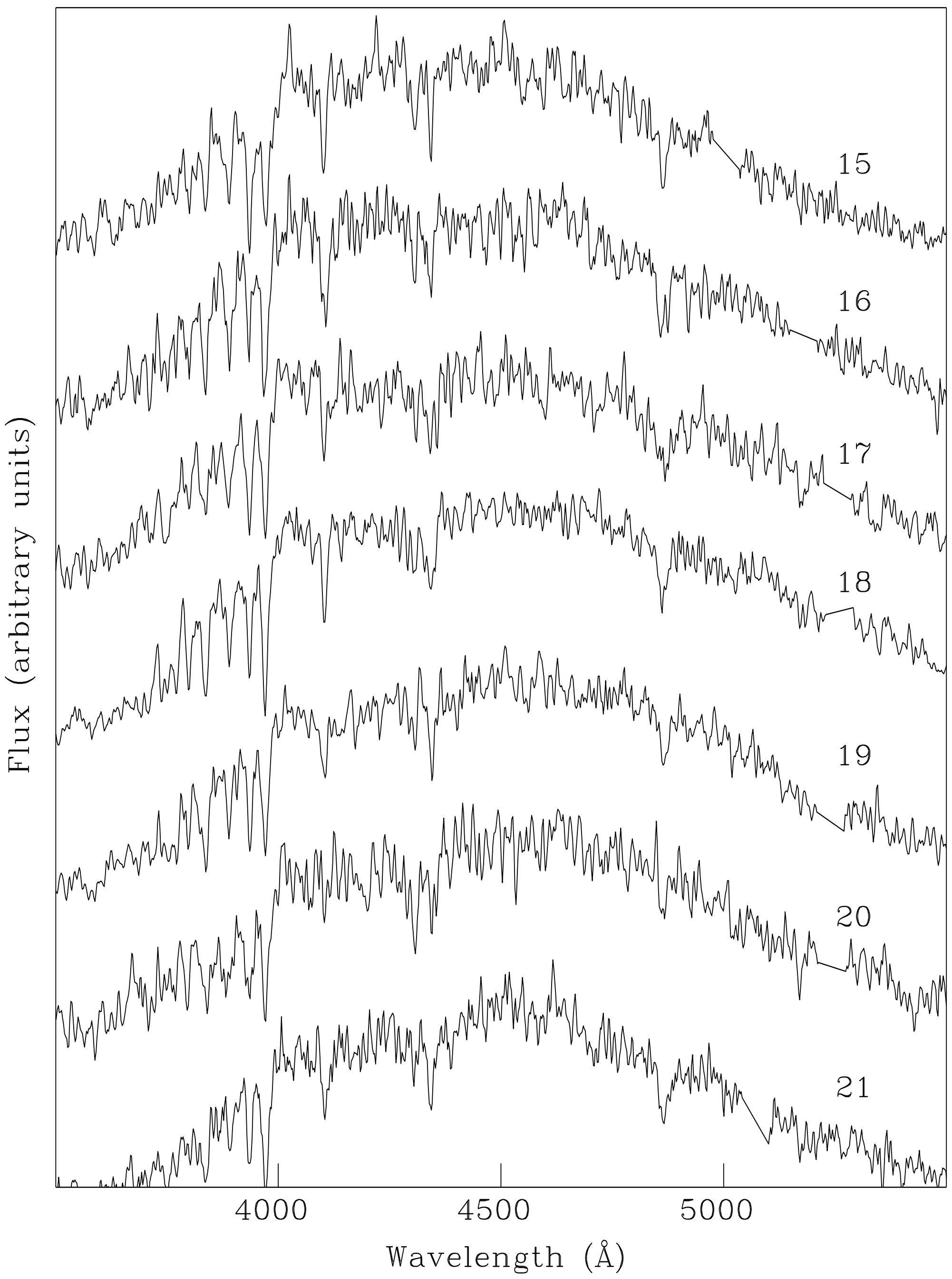

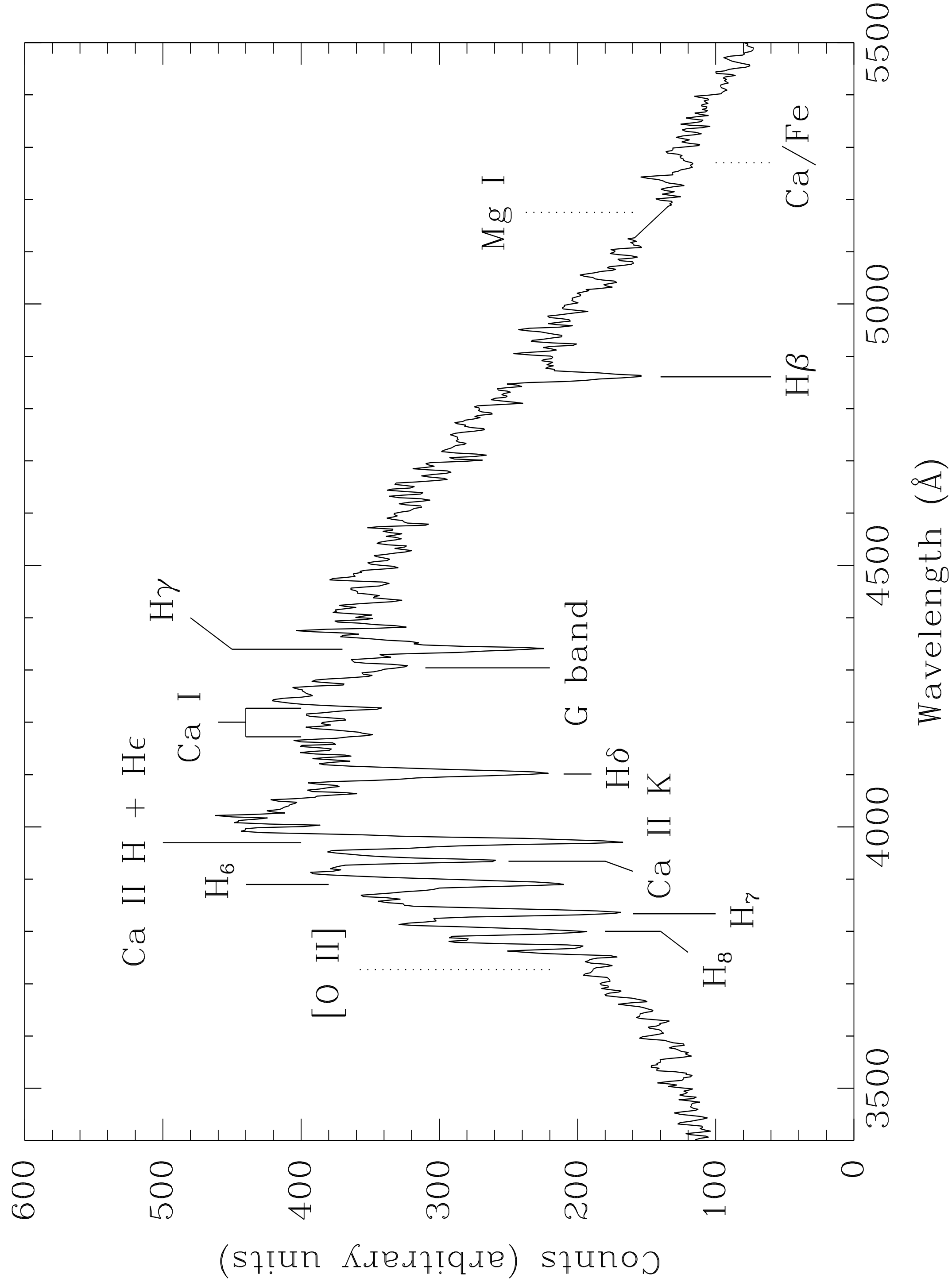

Fig. 2.—

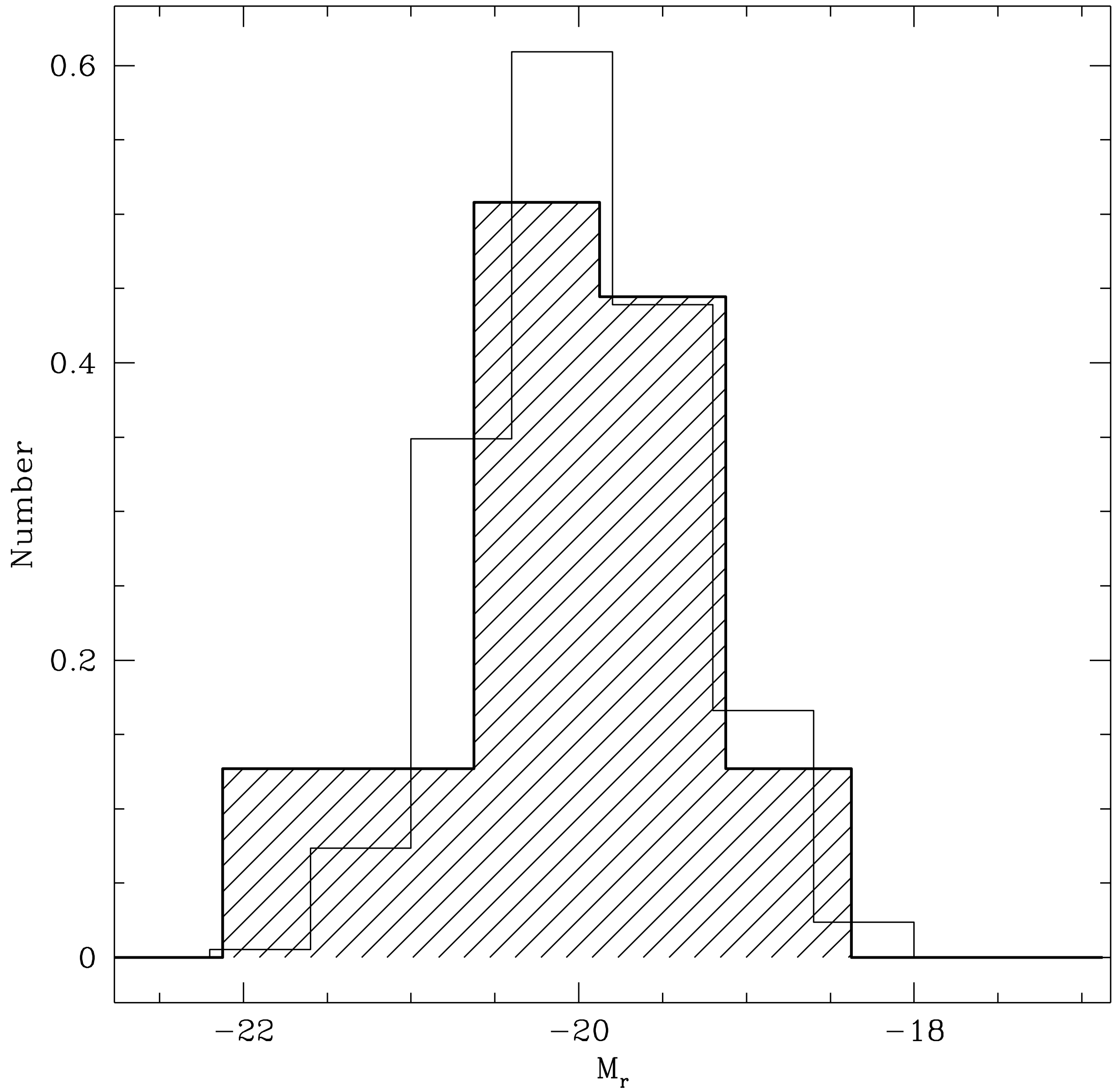

Fig. 3.—

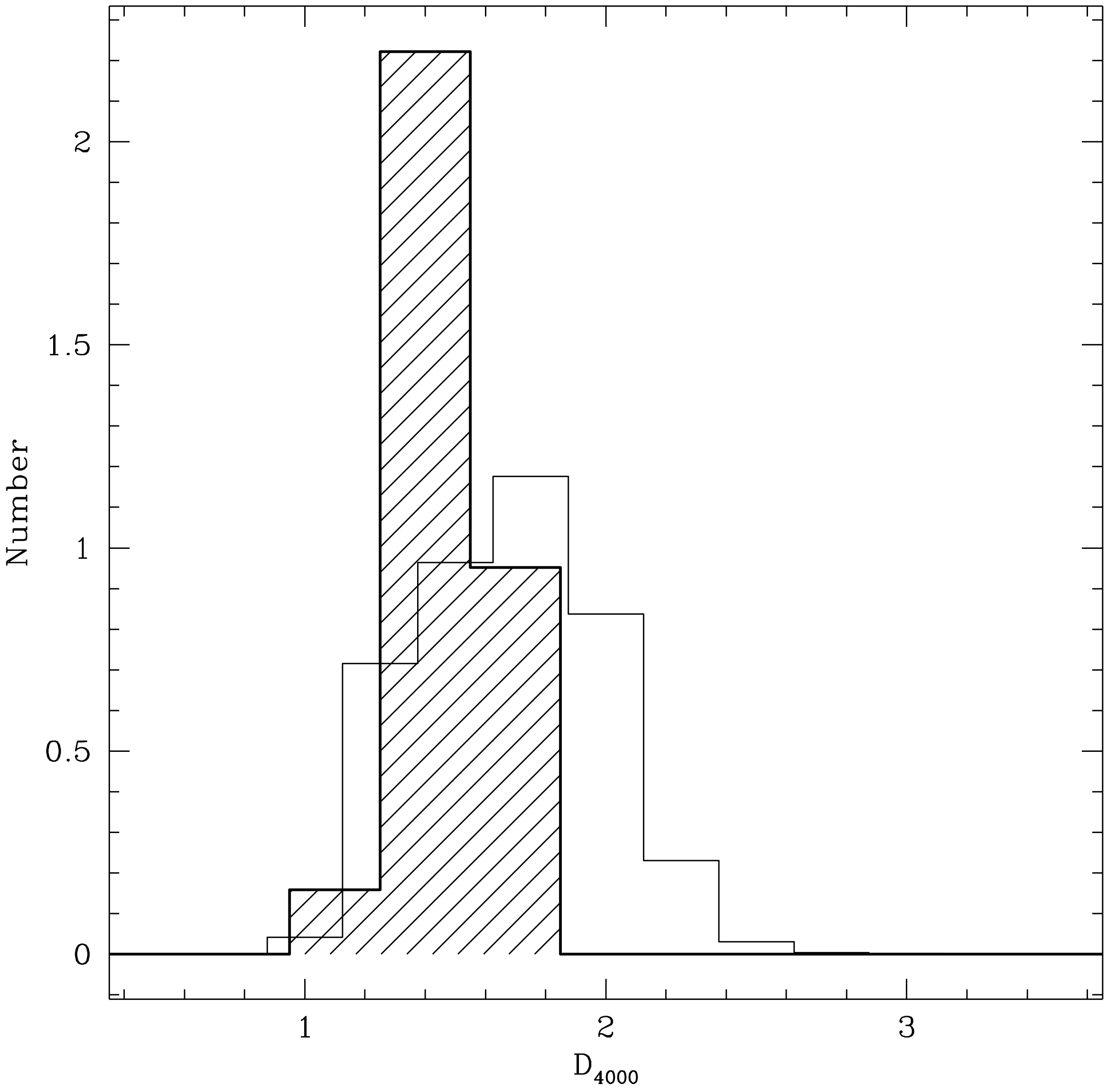

Fig. 4.—

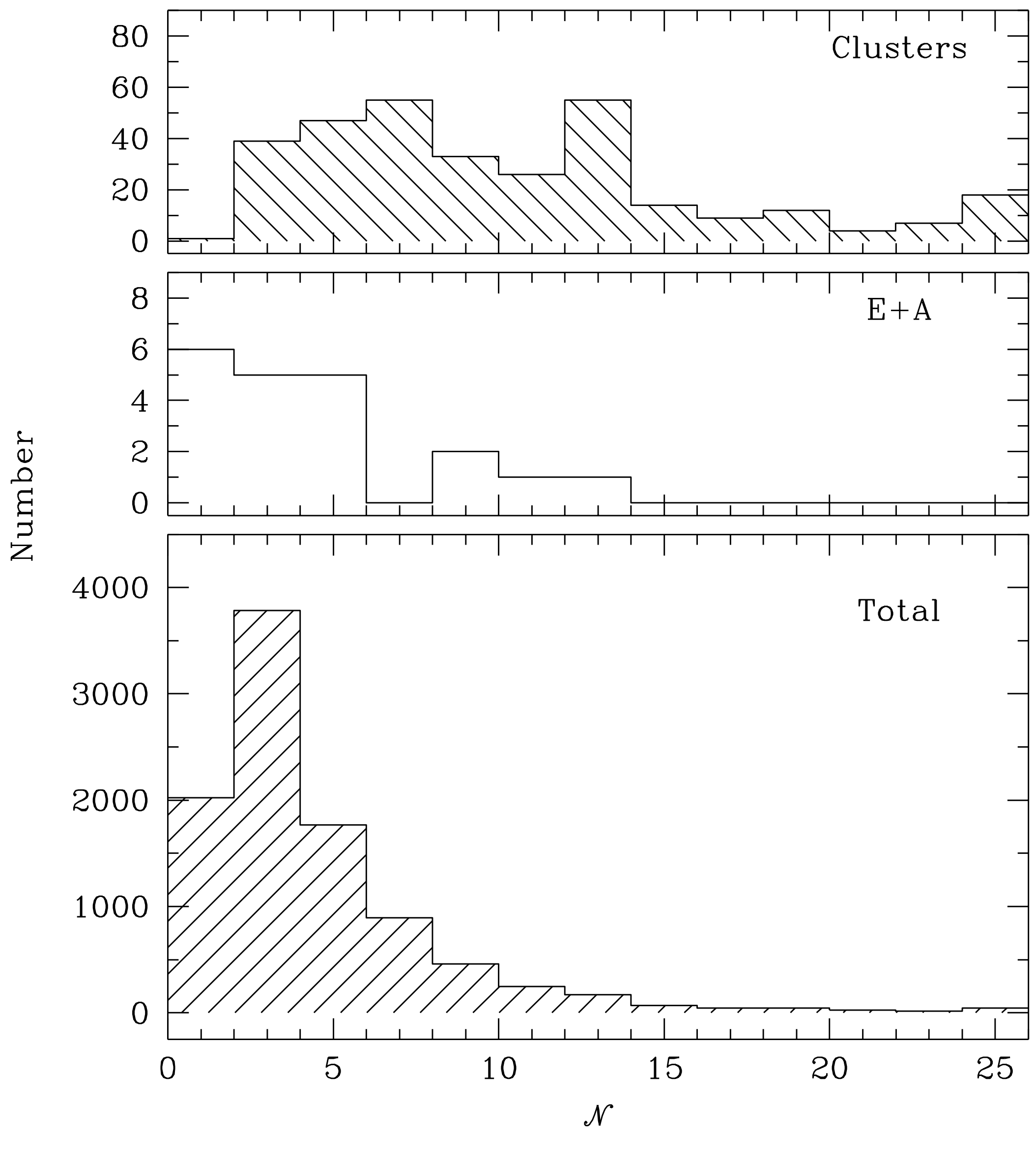

Fig. 5.—

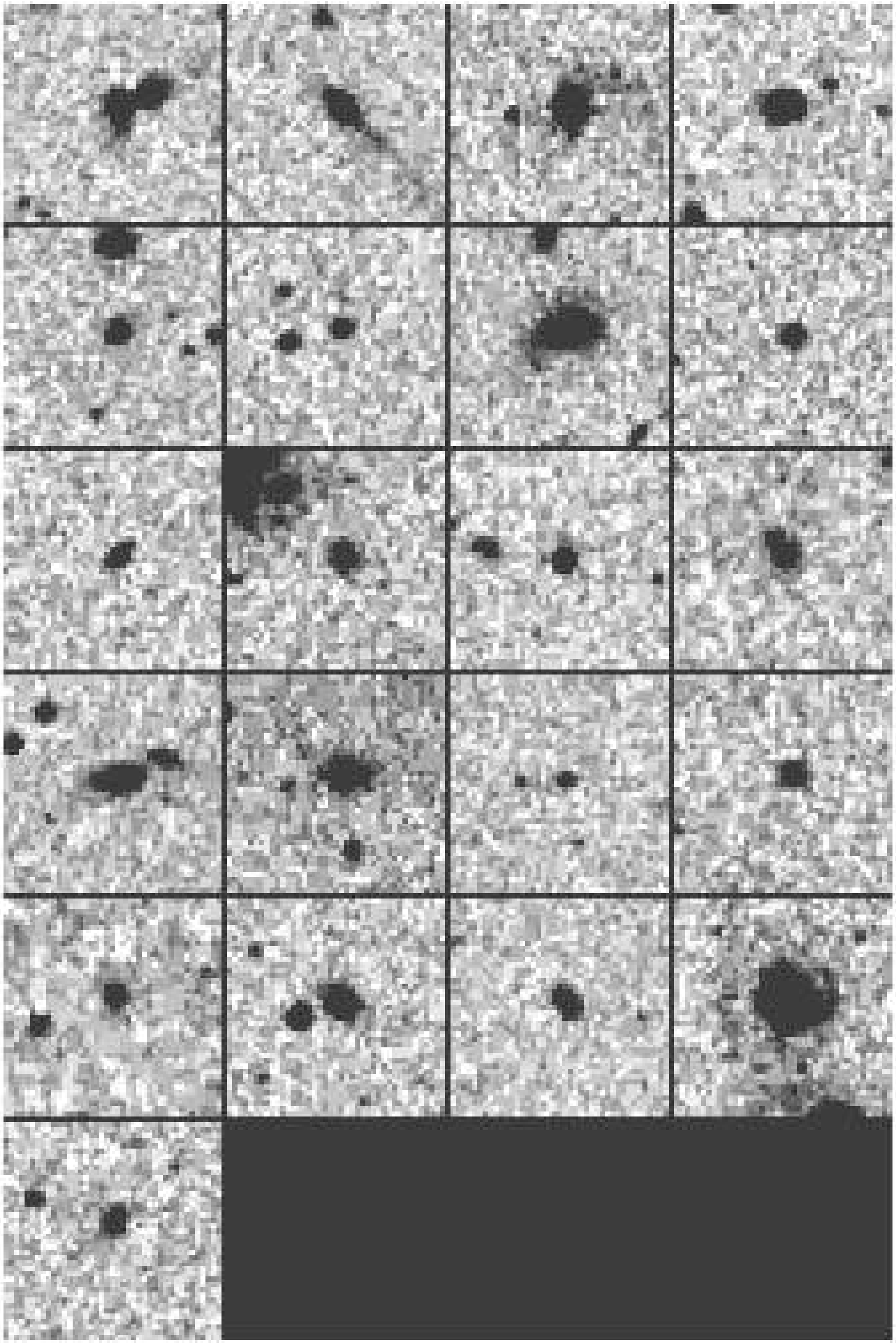